# Nitrogen doping of TiO$_2$ photocatalyst forms a second e$_g$ state in the Oxygen (1s) NEXAFS pre-edge

Artur Braun[1a], Kranthi K. Akurati[1], Giuseppino Fortunato[2], Felix A. Reifler[2], Axel Ritter[2], Ashley S. Harvey[3], Andri Vital[1], Thomas Graule[1,4]

[1] Laboratory for High Performance Ceramics

EMPA – Swiss Federal Laboratories for Materials Testing & Research

CH – 8600 Dübendorf, Switzerland

[2] Laboratory for Advanced Fibers

EMPA – Swiss Federal Laboratories for Materials Testing & Research

CH – 9014 St. Gallen, Switzerland

[3] Department of Materials, Nonmetallic Inorganic Materials

ETH Zurich – Swiss Federal Institute of Technology

CH-8093 Zurich, Switzerland

[4] Technische Universität Bergakademie Freiberg

D-09596 Freiberg, Germany

---

[a] *Corresponding author. Phone +41 (0)44 823 4850, Fax +41 (0)44 823 4150, email: artur.braun@alumni.ethz.ch

**Abstract**

Close inspection of the pre-edge in oxygen near-edge x-ray absorption fine structure spectra of single step, gas phase synthesized titanium oxynitride photocatalysts with 20 nm particle size reveals an additional $e_g$ resonance in the VB that went unnoticed in previous $TiO_2$ anion doping studies. The relative spectral weight of this Ti(3d)-O(2p) hybridized state with respect to and located between the readily established $t_{2g}$ and $e_g$ resonances scales qualitatively with the photocatalytic decomposition power, suggesting that this extra resonance bears co-responsibility for the photocatalytic performance of titanium oxynitrides at visible light wavelengths.

Band-gap tailoring has become an essential step in photocatalysis (PC), for instance as nitrogen doping of $TiO_2$ ($TiO_{2-x}N_x$) in order [1] to make it receptive towards visible light [2]. Improved visible light absorption of $TiO_{(2-x)}N_x$ was originally rationalized by band gap narrowing, caused by mixing of N(2p) with O(2p) states [2]. The correlation between spectral weight of the x-ray photoelectron spectroscopy (XPS) N peak at 396 eV and photocatalytic activity suggests that active sites in $TiO_{2-x}N_x$ are of substitutional nature [2]. Computational studies [3] and valence band (VB) XPS data [4,5] support the suggestion that nitrogen-doping forms new localized states in the band gap just above the VB maximum, leading to substantial reduction of the energy necessary to form oxygen vacancies in bulk $TiO_2$. Recent near edge x-ray absorption fine structure (NEXAFS) spectroscopy studies at the O K-edges and Ti K and $L_{2,3}$ edges [6,7] find that $TiO_{2-x}N_x$ is locally similar to anatase, but point defects represented by oxygen vacancies alter the short range Ti environment, concomitant with tetragonal distortion from octahedral site symmetry [8]. For samples with a large fraction of Ti in five-fold coordination this holds also for the medium-range scale [6]. Here mainly the Ti $e_g$ peak is influenced. In these cases, an important increment of peak intensities in the oxygen edge that corresponds to the antibonding O(2p) transition was observed, probably associated with smaller Ti-O interaction. The $d_{z2}$ and $d_{x2-y2}$ orbitals of the eg states are directed towards ligand anions and therefore more sensitive to deviations from octahedral symmetry [9,10]. Consequently, splitting of $e_g$ states into $d_{x2-y2}$ indicates the degree of distortion from octahedral symmetry. The difference between the crystal field splitting observed in the Ti(2p) and O(1s) NEXAFS spectra arises from the non-symmetric splitting of the $e_g$ states into the $d_{z2}$ and $d_{x2-y2}$ orbitals [8]. The Ti $e_g$ orbitals, directed towards the oxygen, show stronger coupling with the O(2p) orbitals and their antibonding hybridized states are at higher energies than those of the $t_{2g}$ orbitals [9,11].

Two peaks are found in the $e_g$ region of PbCaTi-oxide and PbLaTi-oxide, which are cation doping studies in an $ABO_3$ type perovskite [10,11]. Changes in the oxygen pre-edge region of PbLaTi-oxide were attributed to altered hybridization between O(2p) and Ti(3d) and Pb(6p) orbitals [10]. The oxygen pre-edge region on $TiO_2$ and doped $TiO_2$ is considered as built up from one $t_{2g}$ and one $e_g$ symmetry peak. While we can confirm this for single crystal $TiO_2$, the width and asymmetry of the pre-edge of our $TiO_{2-x}N_x$ require inclusion of a third peak near $e_g$ for a correct deconvolution. No differences in the O and Ti NEXAFs spectra in $TiO_2$ and N, S and C doped $TiO_2$ are apparent in [7], whereas noticeable increase of spectral intensity in the VB XPS near $E_F$.

For one-step gas phase synthesis of $TiO_{(2-x)}N_x$ in combined oxidizing and nitriding atmospheres, a custom-made alumina tube hot wall reactor was fed with $TiCl_4$ and $N_2$ (purity > 99.995%) carrier gas, and with $NH_3$ (purity 99.99%) and water vapor transported by air, all of which controlled by separate mass flow controllers and rotameters, so as to produce at various process temperatures yellowish N-doped $TiO_2$ nanoparticles which were collected on borosilicate glass fiber filters. Process parameters ranged as follows: 600 μl/min < $NH_3$ < 2400 μl/min; $TiCl_4$ = 1.38 g/h; 0.48 g/h < $H_2O$ < 3.12 g/h with 81 ml/min air carrier gas; 600 μl/min < $N_2$ < 2400 μl/min; T = 650°C or 750°C. One potential formation and reaction path for our route is that $NH_3$ and $H_2O$ form $NH_4OH$, which then reacts with $TiCl_4$, which in turn hydrolyzes to form $Ti(OH)_4$, while the Cl reacts with $NH_4OH$ to form $NH_4Cl$. Rinsing the nanopowder in water with subsequent centrifugation yields pure $TiO_{(2-x)}N_x$, the x-ray diffractograms of which could be matched with the anatase reference pattern JCPDS 21-1272. Transmission electron micrographs revealed an average primary particle size of around 20 nm. Soft x-ray absorption spectra were recorded at 300 K and 5 x E-10 Torr vacuum pressure at beamline 1.1 of the Daresbury Laboratory Synchrotron Radiation Source, United Kingdom. The O K-edge and Ti $L_{2,3}$ edge spectra were collected from fine powder, dispersed over conducting carbon tape, with the surface sensitive total electron yield at an energy resolution of $\Delta E/E \sim 1/5000$. A $TiO_2$ (001) single crystal (CRYSTEC, Berlin) spectrum was recorded at beamline 9.3.2 at the Advanced Light Source in Berkeley, California under the same conditions like in Daresbury. Transmission electron microscopy showed spherular primary particles with 20 nm size. Qualitative and quantitative information on the N doping was obtained by XPS (PHI LS5600 equipped with Mg Kα X-ray source, spectra calibrated using C1s = 285.0 eV). Decomposition of nicotine and 3-methyl-2-hexenoic acid (sweat acid) in gaseous phase under controlled relative humidity in a 20 ml vial was carried out under visible light (400 nm – 500 nm wavelength) and quantified with a gas chromatograph. We briefly summarize the influence of process parameters on the doping: Significant nitridation occurs with $NH_3$ flow rate > 1800 μl/min, but is reduced with increasing T. Moisture flow rate > 0.48 g/h yields no nitridation. Based on the decomposition and NEXAFS studies, we could break down all samples into three classes of photocatalysts (PC), ranked as "good", "fair", and "poor". Well performing catalysts (*good*) require generally high moisture content, high synthesis temperature and high $NH_3/N_2$ flow. *Poor* performing materials are synthesized at low moisture, low temperature, or at very high N concentration.

Ti(2p) XPS data show that $Ti^{4+}$ is the dominant species, with minor contributions from $Ti^{3+}$. The N(1s) peak at 396 eV is indicative to Ti-N bonds [2] and particularly large for the *poor* catalyst, and smallest yet noticeable for the *good* catalyst (Figure 1). The decomposition rate can be directly compared with the NEXAFS spectra recorded at the oxygen edge and the Ti $L_2$ and $L_3$ edges.

We begin with the oxygen spectrum of "*good*" PCs. Figure 2-a shows a representative O(1s) NEXAFS spectrum of these samples. The high intensity pre-peak at about 532 eV, for the *good* PC, is split into the [6,7,12] $t_{2g}$ and $e_g$ like peaks that arise from Ti(3d)-O(2p) hybridization. The $t_{2g}$ at 531.28 eV (1.6 eV width) leads in intensity, followed by the $e_g$ peak at 534.27 eV (width 2.3 eV). The Ti $e_g$ orbitals are directed towards the O anions. Their coupling with the O(2p) orbitals is stronger and their antibonding hybridized states are at higher energy than those of the $t_{2g}$ orbitals [9,10]. The sharpness of these two peaks reflects high ionic or covalent bonding between the O and Ti. The energy separation of the $e_g$ and $t_{2g}$ peaks is around 2.7 eV, in line with literature [12]. Since this "*good*" photocatalyst is doped with 0.1 at. % N, hybridization of states from N cannot be ruled out. The O(2p) absorption peak is located at 539 eV. The broad peak at about 544 eV is due to O(2p) and Ti(4sp) hybridized states. Asymmetry and width of the $e_g$ peak suggest that a second peak must be included for fitting in order to proper deconvolute the spectrum, as exercised in Figure 2-a. Comparison with the spectral shape of the pre-peaks in the "*fair*" and "*poor*" samples further justifies inclusion of such second $e_g$ peak in the deconvolution, which we tentatively assign to $e_g\downarrow$. The inset in the spectrum for the *good* photocatalyst shows its magnified pre-edge region and also that of a pure $TiO_2$(001) single crystal, latter of which is deconvoluted and fitted with 2 Voigt functions of 1.5 eV width for the $t_{2g}$ at 531.25 eV and 2.0 eV width for the $e_g$ peak at 533.90 eV. The fit parameters for the *good* PC are 531.28 eV, 1.6 eV width; 532.98 eV, 2.2 eV width; 534.27 eV, 2.3 eV width.

Figure 2-b shows a representative oxygen spectrum of the next best set of catalysts ("*fair*"). Though the pre-edge is still split, the splitting is less sharp than in the spectrum of the "*good*" photocatalyst, Figure 2-a. The height of this peak relative to the O(2p) peak is also smaller, but the O(2p) and O(2p)-Ti(4d) features develop with those in Figure 2-a. While the spectrum in Figure 2-a can hypothetically be fitted with only 2 Voigt functions, provided they were sufficiently broad, this would not work for the spectrum in Figure 2-b.

The representative spectrum of the catalyst with least performance ("*poor*") is shown in Figure 2-c. The $t_{2g}$ symmetry peak persists merely as a shoulder, but we notice a slight shift of intensity at 531 eV towards the Fermi energy $E_F$, when compared with the other spectra. The new $e_g\downarrow$ feature rising up in Figure 2-a at 532.5 eV is relatively well pronounced in Figure 2-c. Differences with respect to the *good* catalyst spectra persist at the O(2p) region at 538 eV and beyond. The O(2p) peak and the peak from the hybridized states of Ti(4sp)-O(2p) seem to merge into one broad peak at about 542 eV. Direct comparison with pure $TiO_2$ spectra demonstrates that N-doping creates a new $e_g\downarrow$ state between the $t_{2g}\uparrow$ and the $e_g\uparrow$ state. Close inspection of the work by [6] shows their $TiO_{2-x}N_x$ oxygen spectra, too, indicate such extra intensity, whereas the spectra for $TiO_{2-x}N_x$ in [7] lack such feature, i.e. are virtually identical with $TiO_2$. For example, hole doping upon lithiation in $Li_{(1+x)}Ti_{(2-x)}O_4$ introduces a new $e_g\uparrow$ peak between the $t_{2g}\uparrow$ peak and the Fermi level [14]. Localized, isolated N(2p) states above the O(2p) VB maximum of $TiO_2$ were predicted [3,14]. Two $e_g$ pre-edge peaks in oxygen NEXAFS spectra on Ti compounds have been found in $ABO_3$ type perovskites [10,11]. Low level N doping creates a large concentration of oxygen vacancies [3,6] and thus noticeable spectral intensity.

In the deconvolved Ti L-edge spectra we have accounted for all known transitions, as exercised for the *good* PC, Figure 2-d [7,15]. Comparison with the spectra in [15] shows this spectrum is reminiscent with anatase, specifically because of the intensity ratio of the two $e_g$ peaks on the $L_3$ edge. The same applies to the spectrum of the *fair*

PC, which virtually not differs from the *good* PC spectrum. Notable differences between the *poor* PC and the two other PC is the broad shoulder-like characteristic of the $L_3$ $t_{2g}$ peak, which is not as well separated from the $L_3$ $e_g$ peaks, and the larger contribution from the two pre-edge peaks causing the shoulder at about 457 eV, which indicate multiplett splitting of electron-hole interaction [15]. These originate from dipole forbidden transitions and are typical for $d^0$ species like $Ti^{4+}$ [9]. This peak has triplet character and is mixed into the $L_3$ edge by spin orbit coupling and Coulomb repulsion [9]. Since this effect is pronounced in the N-rich doped $TiO_2$, we believe the high concentration of oxygen vacancies is the major origin for the peak. Large oxygen vacancy concentration implies a reduction of the nominal $Ti^{4+}$ in $TiO_2$ towards $Ti^{3+}$ upon doping. We indeed observe a slight chemical shift of the spectrum of the *poor* PC, the pre-edge peaks of which are located at 456.9 eV and 457.7 eV, whereas those for the two other PC are located at 457.3 eV and 457.9 eV. Similar chemical shift holds for the $L_3$ and $L_2$ structures, and the aforementioned shift of the oxygen spectrum of the *poor* PC towards $E_F$.

The spectral link with the photocatalytic properties can most likely be made with the VB, observable at the pre-edges of the oxygen spectra. A qualitative overview is presented in Figure 3, where the ratio S of the intensity of the new doped $e_g\downarrow$ peak with the readily existing $e_g\uparrow$ peak and $t_{2g}\uparrow$ peak, which are native to $TiO_2$, is formed: S = $(t_{2g}\uparrow + e_g\uparrow)/e_g\downarrow$. This ratio is large for the *good*, less doped PC, and small for the highly doped, *poor* performing PC. The relative decomposition power of the three catalysts (*good, fair, poor*) is paralleled by the variation of the spectral pre-edge peak ratio S. This phenomenological trend may not be simply generalized, because the highest such ratio is by definition obtained when no extra $e_g\downarrow$ peak is present, i.e. in non-doped $TiO_2$, for which the relative decomposition is 40% for nicotine (24% for sweat acid).

We see, however, that presence of this peak is detrimental to the PC performance. This is in line with the finding that there exists an optimum nitrogen doping level (0.25 at. %, [2]), beyond which the PC performance is less with respect to visible light activity.

Funding by E.U. MIRG # CT-2006-042095, Swiss NSF # 200021-116688, and Swiss Federal Office of Energy # 153613-102809, 153613-102691, and Swiss CTI project NANODOR # 7023.3;4. The ALS is supported by the Director, Office of Science/BES, of the U.S. DoE, # DE-AC02-05CH11231. The Science and Technology Facilities Council of the UK are acknowledged for beamtime at the Synchrotron Radiation Source, project # 47093.

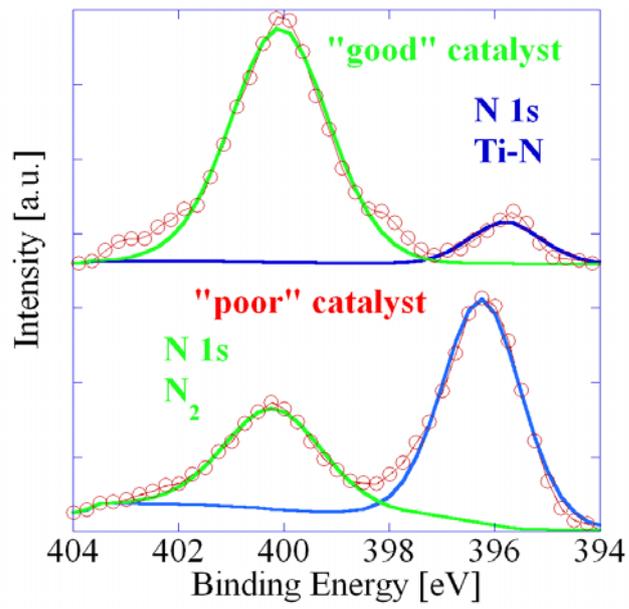

Figure 1: N 1s XPS spectra of N-doped TiO$_2$. Top shows the *good* PC, bottom shows the *poor* PC. The peak at about 396 eV is indicative to Ti-N bonds.

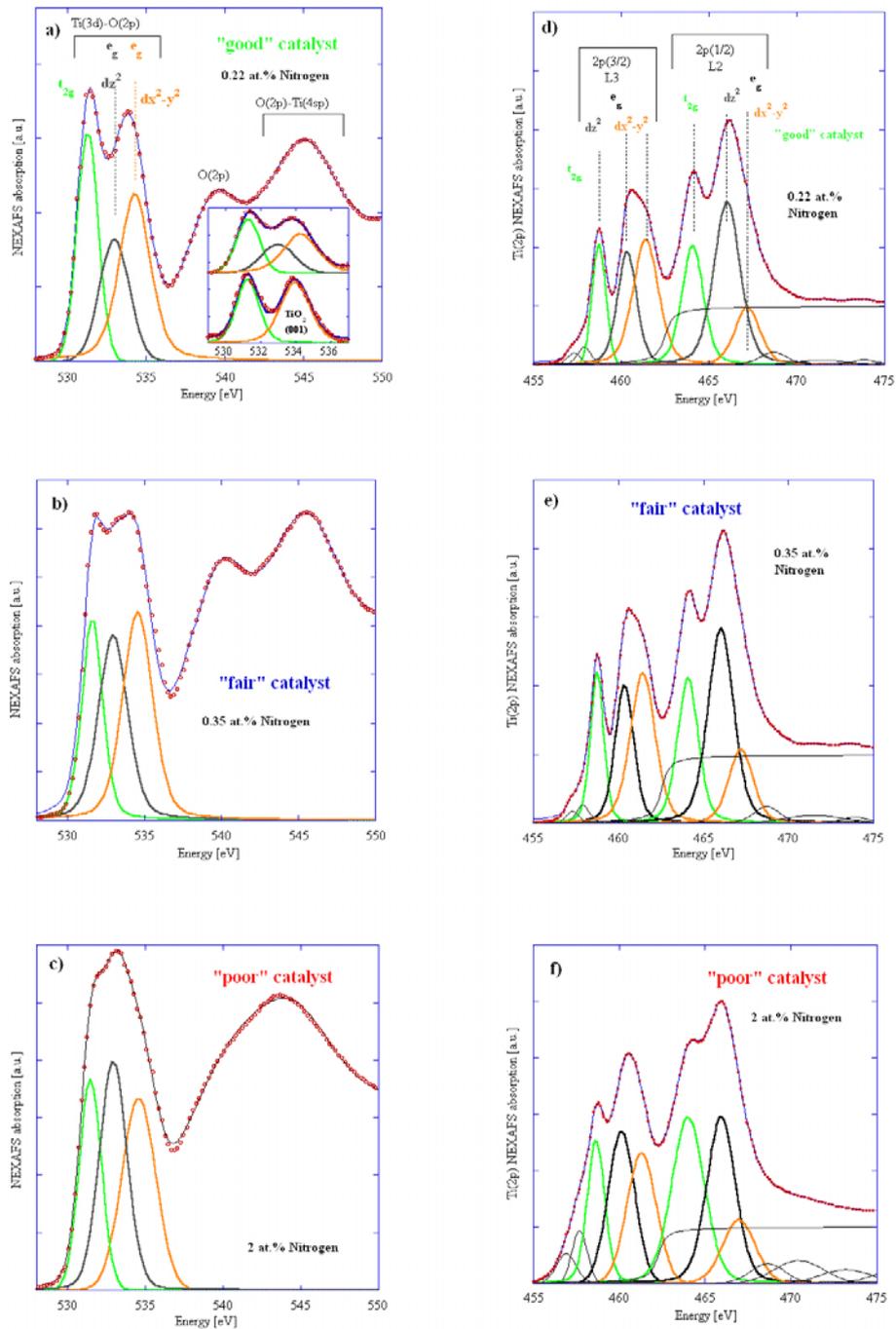

Figure 2: O(1s) (left panel) and Ti(2p) (right panel) NEXAFS spectra (open symbols) with least square fit (solid thin blue line) of Voigt functions and arctan functions. Inset in a) compares pre-edge of *good* PC with $TiO_2(001)$.

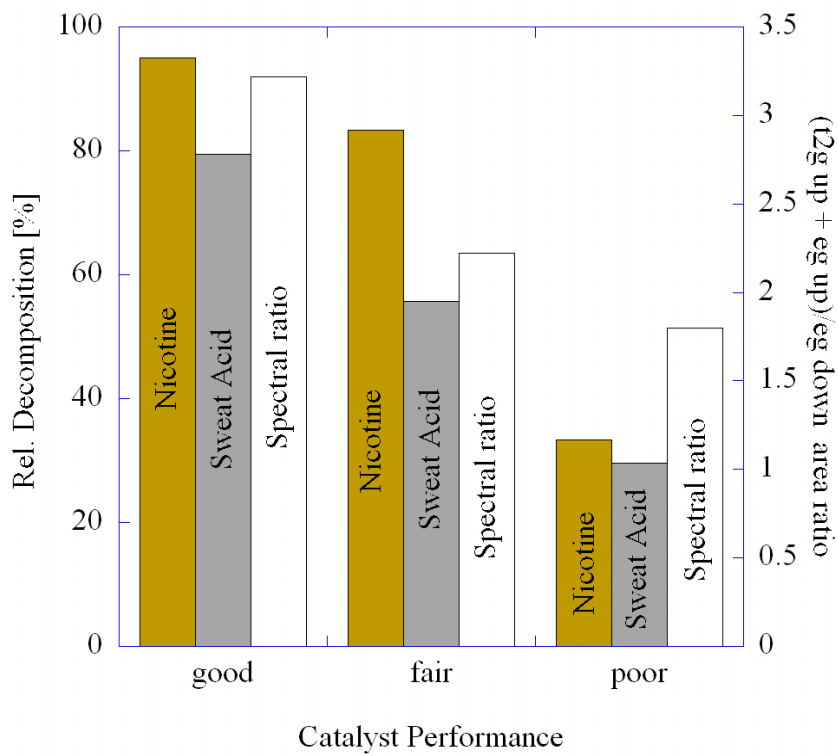

Figure 3: Nicotine and sweat acid decomposition rate in [at. %] compared with the spectral ratio from oxygen NEXAFS (right axis).